\documentclass[11pt,a4paper]{article}
\usepackage{amsfonts,graphicx}

\newtheorem{lem}{Lemma}
\newtheorem{cor}{Corollary}
\newtheorem{prop}{Proposition}
\newtheorem{defn}{Definition}

\newcommand{\cL}{{\cal L}}
\newcommand{\1}{\mathbb{I}}
\newcommand{\U}{{\sf U}}
\newcommand{\J}{{J}}
\newcommand{\Lp}[1]{\Pi_{#1}}
\newcommand{\LG}{\Lambda}
\newcommand{\G}{\Gamma}

\newcommand{\Js}[1]{F_{+{#1}}}
\newcommand{\Jsb}[1]{F_{-{#1}}}
\newcommand{\Jsd}[1]{F_{\pm{#1}}}
\newcommand{\Jsa}[1]{F^{\dagger}_{-{#1}}}
\newcommand{\Jsab}[1]{F^{\dagger}_{+{#1}}}
\newcommand{\Jfd}[1]{{\sf F}_{\pm{#1}}}
\newcommand{\Jfa}[1]{{\sf F}^{\dagger}_{-{#1}}}
\newcommand{\Jfab}[1]{{\sf F}^{\dagger}_{+{#1}}}
\newcommand{\Jfad}[1]{{\sf F}^{\dagger}_{\pm{#1}}}
\newcommand{\Sw}{\Psi}
\newcommand{\eSw}{\Xi \rule{0mm}{4.5mm}}\newcommand{\M}[2]{M^{#1}_{+{#2}}}
\newcommand{\eSwa}{\Xi^{\dagger} \rule{0mm}{4.5mm}}
\newcommand{\Ma}[2]{M^{\dagger\,{#1}}_{-{#2}}}
\newcommand{\Mb}[2]{M^{#1}_{-{#2}}}
\newcommand{\Md}[2]{M^{#1}_{\pm{#2}}}

\newcommand{\js}[1]{f_{+{#1}}}
\newcommand{\jsb}[1]{f_{-{#1}}}

\newcommand{\R}{\mathbb{R}}
\newcommand{\C}{\mathbb{C}}
\newcommand{\T}{\mathbb{T}}

\newcommand{\Ran}{\mbox{Ran}}

\newcommand{\Def}{{\cal N}}
\newcommand{\cG}{{\cal G}}
\newcommand{\cH}{{\cal H}}
\newcommand{\Dom}{\mbox{Dom}}

\title{Hermitian symplectic geometry and extension theory\thanks{AMS: Primary 47A20; Secondary 34B45, 34L40, 47A40, 81U20}}
\author{M. Harmer}
\date{}

\begin{document}

\maketitle 

\begin{abstract}
Here we give brief account of hermitian symplectic spaces, showing that
they are intimately connected to symmetric as well as self-adjoint
extensions of a symmetric operator. Furthermore we find an explicit 
parameterisation of the Lagrange Grassmannian in terms of the unitary
matrices $\U (n)$. This allows us to explicitly describe all self-adjoint
boundary conditions for the Schr\"{o}dinger operator on the graph in
terms of a unitary matrix. We show that the asymptotics of the scattering
matrix can be simply expressed in terms of this unitary matrix. \\     
\end{abstract}

\section{Introduction}

The main motivation to study hermitian symplectic spaces---this
terminology follows \cite{Kost:Sch}---is the well know connection
between the self-adjoint extensions of a symmetric operator and the
Lagrange planes of a  hermitian symplectic space \cite{Pav,Kost:Sch,Nov3}.
This is based on the fact that the boundary form of a symmetric operator
is a hermitian symplectic form and the extensions of the operator may be
identified with isotropic subspaces in the associated hermitian
symplectic space. \\
In the first section of this paper we define and describe some of the
properties of hermitian symplectic spaces. By our definition hermitian
symplectic spaces (unlike symplectic spaces) need not be even dimensional
or admit a canonical basis. We show that when a hermitian symplectic
space admits a canonical basis, it has Lagrange planes and derive an
explicit parameterisation of the set of Lagrange planes in terms of the
set of unitary matrices $\U (n)$ where $n$ is half the dimension of the
space. \\ 
In the following section we consider connections to extension
theory of symmetric operators. It is observed that hermitian symplectic
spaces that do not admit a canonical basis, or Lagrange planes,
correspond to symmetric operators with unequal deficiency indices (in
this case the extensions are described by isotropic subspaces). On the
other hand, symmetric operators with equal deficiency indices correspond
to hermitian symplectic spaces with Lagrange planes and as is well known
these Lagrange planes may be used to describe the self-adjoint
extensions. The fact that the set of Lagrange planes, or self-adjoint
extensions, is isomorphic to $\U (n)$ is in accordance with the
parameterisation of self-adjoint extensions by a unitary map between the
deficiency subspaces as described by Neumann extension theory
\cite{Akh:Glz}. \\  
We then consider the specific example of the
Schr\"{o}dinger operator on the graph. Our explicit parameterisation of
the Lagrange planes in terms of the unitary matrices allows us to
describe all self-adjoint boundary conditions at the origin for the
Schr\"{o}dinger operator on the graph with trivial compact part in terms
of a unitary matrix. Furthermore, we show that the asymptotics of the
scattering matrix may be written in terms of this unitary matrix and that
the boundary conditions do not contribute to the discrete spectrum iff
this unitary matrix is also hermitian. We also use a property of this
parameterisation, as well as the Wronskian, to show the unitarity of the
scattering matrix. \\ 

\section{Hermitian symplectic geometry}

Many of the basic ideas in this section can be found in any standard text 
on symplectic geometry \cite{Arn,Arn:Giv,Fom,Mar:Rat}. However, the
concept of a canonical hermitian symplectic space and the details of the
parameterisation of Lagrange planes in a hermitian symplectic space
distinguish this construction from the standard symplectic case. In
particular the Lagrange planes in hermitian symplectic geometry are
parameterised by unitary matrices whereas they have different
parameterisations in the standard symplectic geometry. Also, by our
definition, a hermitian symplectic space need not be even dimensional or
admit a canonical basis---unlike the symplectic case. This is
seen to correspond to a symmetric operator with unequal deficiency
indices. \\
\begin{defn}
The two-form $\langle\cdot,\cdot\rangle$, linear in the second
argument and conjugate linear in the first argument, is a hermitian
symplectic form if
\[
\langle\phi,\psi\rangle=-\overline{\langle\psi,\phi\rangle} .
\]
\end{defn}
We recall that the standard symplectic form obeys
$\langle\phi,\psi\rangle=-\langle\psi,\phi\rangle$. We will use the
prefix `hermitian' to emphasise this distinction.
\begin{defn}
We say that an m-dimensional ($m <\infty$) vector space $H_{m}$ over $\C$
is a hermitian symplectic space if it has defined on it a
nondegenerate hermitian symplectic form. By nondegenerate we mean that if
$\phi$ obeys
\[
\langle\phi,\psi\rangle=0 \qquad \forall \psi\in H_{m}
\]
then $\phi=0$.
\end{defn}
Since $H_{m}$ is a vector space we can find a basis $\{ e_i \}^{m}_{i=1}$
for it and use this basis to express the hermitian symplectic
form as a matrix with entries
\begin{equation}\label{sym4} 
\omega_{ij} = \langle e_i , e_j \rangle .
\end{equation}
By the definition of the form, the matrix $\omega$ is a skew-hermitian, 
$\omega = -\omega^{\star}$, nondegenerate matrix. Clearly the hermitian 
symplectic form can be written
\begin{equation}\label{sym3} 
\langle\phi,\psi\rangle=(\phi ,\omega \psi )
\end{equation}
where, on the right hand side, $\phi$ and $\psi$ are written as vectors in
$\C^{m}$ using the basis $\{ e_i \}^{m}_{i=1}$ and $(\cdot ,\cdot)$ is the
standard hermitian scalar product on $\C^{m}$, making it an
$m$-dimensional Hilbert space. \\ 
In the usual symplectic case $\omega$ is
skew-symmetric and hence, due to nondegeneracy, of even order. This
restriction does not apply to skew-hermitian matrices and hence there is
no obstruction to having hermitian symplectic spaces of odd dimension. \\
Hermitian symplectic
spaces differ from symplectic spaces in another important respect; given
any symplectic space it is always possible to find a {\em canonical
basis:}
\begin{defn}
A basis $\{ p_{i},q_{i}\}^{n}_{i=1}$ which has the following property
\begin{eqnarray*}
\langle p_{i},q_{j}\rangle= \delta_{ij} 
=-\langle q_{j},p_{i}\rangle \\
\langle p_{i},p_{j}\rangle= 0 
=\langle q_{i},q_{j}\rangle 
\end{eqnarray*}
where $\delta_{ij}$ is the Kronecker delta is known as a canonical
basis.
\end{defn}
Even an even-dimensional hermitian symplectic space, $H_{2n}$, need not
admit a canonical basis. Let us suppose that $H_{2n}$ has a basis $\{ e_i
\}^{2n}_{i=1}$ so that the skew-hermitian matrix
$\omega$ is
\[
\omega =\left( \langle e_i , e_j \rangle \right) = i \1_{(2n)} .
\]
We denote by $\1$ or $\1_{(n)}$ the $n\times n$
unit matrix. This case is obviously prohibited in the symplectic
case but acceptable in the hermitian symplectic case. Now if it were
possible to find a canonical basis in this space then there would be a
non-singular transformation of the basis, $P$, such that $\omega$ would
be transformed to
\[
P^{\star} \omega P = \J
\]
where $\J$, known as the canonical symplectic structure, is
\[
\J=\left( \begin{array}{cc}
 0  & \1 \\
 -\1 &  0
\end{array}
\right) .
\]
This is clearly not possible. We use the fact that any
non-singular transformation can be written as the product of a unitary
and a hermitian matrix, $P= U H$. Consequently
\[
i H^2 = P^{\star} \omega P = J
\]
and the left hand side is a matrix with eigenvalues only on the imaginary
axis in the upper half plane. The right hand side, $\J$, however, has
eigenvalues $\pm i$ equally distributed between the upper and lower half
planes. 
\begin{defn}
We say that a hermitian symplectic space is canonical if it admits a
canonical basis.
\end{defn}
In the following we denote
\[
\1_{(n_+ ,n_- )}\equiv\left( \begin{array}{cc}
 \1_{(n_+ )}  &    0  \\
   0     & -\1_{(n_- )}
\end{array}
\right) .
\]
\begin{lem}\label{charhs}
A hermitian symplectic space $H_m$ is, up to a non-singular
transformation of the basis, completely characterised by two integers,
$n_+$, $n_-$, $n_+ + n_- =m$. Specifically the matrix $\omega$
associated with the hermitian symplectic form can be
diagonalised to 
\[
i \1_{(n_+ ,n_-)} .
\]
Furthermore $H_m$ is canonical iff $n_+ =n_-$.
\end{lem}
{\it Proof:} A hermitian symplectic space is
specified by the matrix $\omega$ up to a non-singular
transformation of the basis, $P$. The matrix $-i\omega$ is hermitian and
hence it can be diagonalised
\[
-i\omega = U D U^{\star}
\]
where $D$ is a real diagonal matrix without zeroes on the diagonal. Let
us choose the matrix $H$ as the {\em positive} diagonal matrix so that
$D^2 = H^4$. Then choosing the  non-singular transformation of the basis,
$P= U H^{-1}$ we get 
\[
P^{\star} \omega P = i H^{-1} U^{\star} U D U^{\star} U H^{-1} = i
\1_{(n_+ ,n_- )} 
\]
where $n_{\pm}$ are the number of positive and negative eigenvalues of
$-i\omega$ respectively. Clearly, when $n_+ =n_- =n$ we can
find a canonical basis since we can transform $i \1_{(n,n)}$ to
$\J$. \hspace*{\fill} $\Box$ \\
\begin{defn}
We say that $\phi,\psi\in H_{m}$ are skew-orthogonal, denoted
$\phi\perp\psi$, if
\[  
\langle\phi,\psi\rangle=0 .
\]
\end{defn}
\begin{defn}
Given a subspace $N\subset H_{m}$, we define the skew-orthogonal
complement, $N^{\perp}$, as the subspace
\[  
N^{\perp}\equiv\{\phi; \;\phi\in H_{m},\,
\langle\phi,\psi\rangle=0 \; \forall \psi\in N  \} .
\]
\end{defn}
\begin{defn}
The subspace $N\subset H_{m}$ is isotropic if
\[  
N\subset N^{\perp} .
\]
\end{defn}
Let us assume that we have fixed some basis and found
the corresponding skew-hermitian matrix $\omega$ from equation
(\ref{sym4}) so that $H_m$ can be identified with the Hilbert space
$\C^m$ equipped with a hermitian symplectic form. The remaining lemmata in
this section all have analagous statements in symplectic geometry
\cite{Fom,Mar:Rat}.
\begin{lem}
The subspace $N\subset H_{m}$ is isotropic iff the subspaces
$N$ and
$\omega N$ are orthogonal in $\C^m$.
\end{lem}
{\it Proof:} Follows directly from equation (\ref{sym3}). \hspace*{\fill}
$\Box$ \\
\begin{lem}
The dimension, $k$, of an isotropic subspace $N\subset H_{m}$ never
excedes $m/2$.
\end{lem}
{\it Proof:} Since the operator $\omega$ on $\C^{m}$ is nondegenerate,
the dimensions of $N$ and $\omega N$ are the same. Consequently $k+k\leq
m$. \hspace*{\fill} $\Box$ \\
\begin{defn}
An isotropic subspace $\Lp{n}\subset H_{2n}$ of maximal dimension, that
is dimension $n$, is called a Lagrange\index{Lp} plane.
\end{defn}
\begin{cor}
If $\Lp{n}\subset H_{2n}$ is a Lagrange plane then
$\Lp{n}^{\perp}=\Lp{n}$.
\end{cor}
{\it Proof:}  $\Lp{n}$ and $\Lp{n}^{\perp}$ both have dimension
$n$ and $\Lp{n}\subset \Lp{n}^{\perp}$. \hspace*{\fill} $\Box$ \\

From the definition it is clear that Lagrange planes only exist in
even-dimensional hermitian symplectic spaces, in fact it is not difficult
to show that a hermitian symplectic space contains a Lagrange plane iff
it is canonical. First we need the basic lemma,
\begin{lem}\label{decomp}
Given a hermitian symplectic subspace $V\subset H_{m}$, 
$V^{\perp}$ is also hermitian symplectic,
\[
V + V^{\perp} = H_{m} 
\]
and these subspaces have trivial intersection.
\end{lem}
{\it Proof:} It is clear that the intersection $V\cap V^{\perp}$ is empty.
Supposing instead that there is a $v\in V\cap V^{\perp}$ then $v$ is
skew-orthogonal to all the elements of $V$ and hence the form is
degenerate on $V$ which is a contradiction. \\
Since the matrix $\omega_{ij}$ is nondegenerate the dimension of
$V^{\perp}$ is the codimension of $V$. But since these two spaces do not
intersect, by a simple argument of linear independence
\[
V + V^{\perp} = H_{m} .
\]
Now we suppose that the form is degenerate on $V^{\perp}$, so there is
some element $z\in V^{\perp}$ so that
\[
\langle z , u \rangle =0, \qquad \forall u\in V^{\perp}
\]
and
\[
\langle z , v \rangle =0, \qquad \forall v\in V .
\]
But this would imply that the form is degenerate on $H_{m}$ which is a
contradiction. \hspace*{\fill} $\Box$ \\
\begin{lem}\label{cbeLp}
An even-dimensional hermitian symplectic space $H_{2n}$ is canonical iff
it contains a Lagrange plane.
\end{lem}
{\it Proof:}  It is clear that a canonical hermitian symplectic space
contains a Lagrange plane, viz. the span of the first $n$ elements of the
canonical basis. \\
We suppose that we have an even-dimensional hermitian symplectic space
$H_{2n}$ which contains a Lagrange plane $\Lp{n}$. Then we can find some
basis $\{ e_i \}^{2n}_{i=1}$ so that the first $n$ elements span
$\Lp{n}$. Let us pick $p_1 = e_1$. Since the form is nondegenerate,
there is an element $\hat{q}_1\not\in\Lp{n}$ such that $\langle p_1 ,
\hat{q}_1 \rangle \neq 0$ and hence we can normalise so that
\[
\langle p_1 , q_1 \rangle = 1 .
\]
We denote by $V_1$ the linear span of $\{ p_1 , q_1 \}$. Using the fact
that $\langle p_1 , p_1 \rangle = 0$ it is not difficult to see that
$V_1$ is a canonical hermitian symplectic space. \\
Applying lemma \ref{decomp} to $V_1$ we see that
$V^{\perp}_1$ is a hermitian symplectic space. Furthermore it has a
Lagrange plane given by the span of $\{ e_i \}^{n}_{i=2}$. Repeating this
process for $V^{\perp}_1$ allows us to construct a canonical basis for
$H_{2n}$. \hspace*{\fill} $\Box$ \\
\begin{defn}
A linear transformation is called $\J$-unitary or hermitian symplectic
if it satisfies
\[
g^{\star}\J g=\J .
\]
\end{defn}
Clearly such a transformation takes Lagrange planes to Lagrange planes.
Consider the set of all Lagrange planes of a canonical hermitian
symplectic space $H_{2n}$, the Lagrange
Grassmannian denoted $\LG_{n}$. We show that the Lagrange
Grassmannian is isomorphic to the set of unitary matrices. 
\begin{lem}\label{Can6}
A given Lagrange plane $\Lp{0,n}$ can be made to coincide with
any other Lagrange plane $\Lp{n}$ by means of a hermitian symplectic
transformation of the form
\begin{equation}\label{hs1}
g=\left( \begin{array}{cc}
 A & B \\
-B & A
\end{array}
\right) \hspace{5mm} A,B\in \C^{n\times n}
\end{equation}
where $A$ and $B$ satisfy
\begin{eqnarray}
A^{\star} A + B^{\star} B = \1 \label{kri}\\
A^{\star} B =  B^{\star} A . \label{krii}
\end{eqnarray}
Specifically, if we are given a canonical basis
$\{\xi_{0,i}\}^{2n}_{i=1}$, the first $n$ elements of which span the
Lagrange plane $\Lp{0,n}$, then there is a hermitian symplectic
transformation $g$ such that the first $n$ elements of the canonical
basis $\{\xi_{i}\}^{2n}_{i=1}$ given by 
\[
\xi_{i} = \sum^{2n}_{j=1} g_{ij} \xi_{0,j}
\]
span $\Lp{n}$.
\end{lem}
{\it Proof:} As we are dealing with canonical spaces there always exists
a canonical basis $\{\xi_{0,i}\}^{2n}_{i=1}$ and we choose
$\Lp{0,n}$ to be the span of the first $n$ elements of this basis. In
terms of this canonical basis we can identify $H_{2n}$ with
$\C^{2n}$ where the two-form is given by $\omega =\J$. \\
Consider another arbitrary Lagrange plane $\Lp{n}$. Using the
above identification, $\Lp{n}$ may be considered to be an
$n$-dimensional subspace of $\C^{2n}$. Consequently, we
can find a set of $n$ orthonormal vectors in $\C^{2n}$ which form
a basis for $\Lp{n}$---we denote this basis by $\{\xi_{i}\}^{n}_{i=1}$.
Since the
$\{\xi_{0,i}\}$ form a basis for $H_{2n}$ there are matrices $A$ and $B$
such that
\begin{equation}\label{st1}
\xi_{i}=\sum^{n}_{j}A_{ij}\xi_{0,j}+\sum^{n}_{j}B_{ij}\xi_{0,j+n} 
\qquad \mbox{for} \;i=1,\ldots ,n .
\end{equation}
That is $A_{ij}=(\xi_i ,\xi_{0,j})$, $B_{ij}=(\xi_i ,\xi_{0,n+j})$ for
$j=1,\ldots ,n$. Furthermore, since we have assumed that the $\{\xi_{i}\}$
are orthonormal in $\C^{2n}$ we immediately have 
equation (\ref{kri}). Using the fact that the $\{\xi_{i}\}$ form a
Lagrange plane in equation (\ref{sym3}) gives us 
equation (\ref{krii}). Together these two equations imply that $g$ is a
hermitian symplectic transformation. \hspace*{\fill} $\Box$ \\

In fact, it is easy to see that equations (\ref{kri},\ref{krii}) imply
that $g$ is a hermitian symplectic matrix as well as a unitary matrix,
ie. it preserves the hermitian symplectic form as well as the scalar
product in $\C^{2n}$.
\\  Let us denote by $\cG$ the set of matrices of the form 
\[
\cG=\left\{ g=\left( \begin{array}{cc}
 A  & B \\
 -B & A
\end{array}
\right); \; A,B\in \C^{n\times n}, \; g\in\U (2n) \right\}
\]
which occur in the above lemma, this set is clearly a group under matrix
multiplication. In order to classify $\LG_{n}$ we need to find the      
stationary subgroup of $\cG$, ie.  $\cH\subset\cG$ the elements of which
take the Lagrange plane $\Lp{0,n}$ into itself. But it is easy to see that
in the notation of the above lemma these are just those matrices with   
$B=0$: the stationary subgroup $\cH$ is therefore the set of matrices
\[
\cH=\left\{ h=\left( \begin{array}{cc}
 C & 0 \\
 0 & C
\end{array}
\right); \;  C\in \C^{n\times n}, \; h\in\U (2n) \right\} .
\]
\begin{lem}
The Lagrange Grassmannian $\LG_{n}$ is in one-to-one
correspondence with the unitary group.
\[
\LG_{n}\simeq\cG / \cH\simeq\U (n)
\]
\end{lem}
{\it Proof:} The first isomorphism follows from lemma 6. To see the
second isomorphism we use the unitary matrix
\[
W=\frac{1}{\sqrt{2}}\left( \begin{array}{cc}
 \1  & i\1 \\
 i\1  & \1
\end{array}
\right)
\]
Our choice of $W$ is motivated by the fact that it diagonalises in the
`blockwise' sense matrices of the form given by equation (\ref{hs1}). 
Precisely
\[
WgW^{\star}  =  W \left( \begin{array}{cc}
 A  & B \\
 -B & A
\end{array} \right) W^{\star} 
 =  \left( \begin{array}{cc}
 A-iB  & 0 \\
 0 & A+iB
\end{array} \right) .
\]
Since $g$ is unitary so is $WgW^{\star}$ and hence,
$A-iB$ and $A+iB$ must also be unitary. \\
Now instead of considering the groups $\cG$ and $\cH$, we consider the
unitarily equivalent groups 
\[
\hat{\cG}=W\cG W^{\star}=\left\{ \hat{g}=\left( \begin{array}{cc}
 S & 0 \\
 0 & T
\end{array}
\right); \;  S,T\in\U (n) \right\}
\]
and, since the elements of $\cH$ are already in block diagonal form,
\[
\hat{\cH}=W\cH W^{\star}=\left\{ \hat{h}=\left( \begin{array}{cc}
 C & 0 \\
 0 & C
\end{array}
\right); \;  C\in\U (n) \right\} .
\]
It is easy to see that we can represent the set of cosets
$\hat{\cG} / \hat{\cH}$ by the subgroup of $\hat{\cG}$ consisting of
matrices where  the bottom right block is of the form $T=\1$, that is
\[
\LG_n\simeq\hat{\cG} / \hat{\cH}\simeq
\left\{ \hat{g}=\left( \begin{array}{cc}
 U & 0 \\
 0 & \1
\end{array}
\right); \;  U\in\U (n) \right\} .
\]
This gives the result. \hspace*{\fill} $\Box$ \\
\begin{cor}\label{LPcorr}
A given Lagrange plane can be made to coincide with any other
Lagrange plane by means of a hermitian symplectic transformation of the
form
\begin{equation}\label{hs2}
g=W^{\star}\hat{g}W=W^{\star}\left( \begin{array}{cc}
 U & 0 \\
 0 & \1
\end{array}
\right) W=\frac{1}{2}\left( \begin{array}{cc}
 U+\1 &  i(U-\1) \\
 -i(U-\1) &  U+\1
\end{array}
\right)
\end{equation}
where $U$ is a unitary matrix.
\end{cor}

\section{Extension theory}

Here we consider the extension theory for a symmetric operator $\cL_0$
on a Hilbert space \cite{Akh:Glz, Alb:Kur, Pav, Ree:Sim}. First we
recall some well known facts from operator theory. The domain of the
adjoint operator          
$\cL^{\star}_0$ can be expressed
\[
\Dom (\cL^{\star}_0 )= \Dom (\cL_0 ) + \Def_{+i} + \Def_{-i} 
\]
where these three subspaces are linearly independent. The eigenspaces 
\[
\Def_{\pm i}\equiv \ker (\cL^{\star}_0 \pm i)
\]
are known as the deficiency subspaces and the deficiency indices $(n_+
,n_-)$ are the dimensions of the deficiency subspaces $n_{\pm} \equiv \dim
\Def_{\pm i}$. In what follows we assume $n_{\pm}<\infty$. \\
Typically, the extensions of $\cL_0$ are specified by a unitary map
between the deficiency subspaces \cite{Akh:Glz, Ree:Sim} and self-adjoint
extensions of $\cL_0$ exist when $n_+ =n_-$. Alternatively,
extensions may be described by consideration of the boundary form
\begin{equation}\label{bf0}
{\cal J}(f,g)\equiv (\cL^{\star}_{0}f,g) - (f,\cL^{\star}_{0}g) , 
\end{equation}
where $f,g\in \Dom (\cL^{\star}_{0})$---see \cite{Pav} for a detailed
account. The boundary form ${\cal J}(\cdot ,\cdot)$ is actually a
hermitian symplectic form and when restricted to $\Def_{+i} + \Def_{-i}$ is
nondegenerate, defining a hermitian symplectic space (the form is degenerate
on $\Dom(\cL_0)$, a simple consequence of the fact that $\cL_0$ is
symmetric). 
\begin{prop}
The hermitian symplectic space formed by the boundary form
${\cal J}$ on $\Def_{+i} + \Def_{-i}$ is characterised, in the sense of
lemma \ref{charhs}, by the deficiency indices $n_{\pm}$.
\end{prop}
{\it Proof:} Suppose that we have orthonormal bases 
$\{ \js{,i}\}^{n_+}_{i=1}$, $\{ \jsb{,i}\}^{n_-}_{i=1}$ for $\Def_{+i}$ and
$\Def_{-i}$ respectively. We use these bases to write the boundary form as
a matrix
\[
\omega_{ij} = {\cal J}( \jsb{,i}, \jsb{,j} ) = - 2 i \delta_{ij} .
\]
This completes the proof. \hspace*{\fill} $\Box$ \\

In terms of this hermitian symplectic space it is not difficult to see
that the extensions of $\cL_0$ correspond to isotropic subspaces and,
when the space is canonical (ie. $n_+ =n_-$), that the self-adjoint
extensions correspond to Lagrange planes.

\subsection{The Schr\"{o}dinger operator on the graph with trivial compact
part}

Here we consider the non-compact graph consisting of $n$ semi-axes
connected at a single vertex, we denote such a graph by $\G_n$. 
Functions on $\G_n$ may be represented by elements of the Hilbert space
\[
H(\G_n)=\oplus^{n}_{i=1}L^{2}([0,\infty)) .
\]
The elements of $H(\G_n)$ are $n$-dimensional vector functions and
the inner product on $H(\G_n)$ is
\[
(\phi,\psi) = \sum^{n}_{i=1}(\phi_i,\psi_i)_{L^{2}([0,\infty))} 
 =  \sum^{n}_{i=1} \int^{\infty}_0 \bar{\phi}_i (x) \psi_i (x) dx
\]
where $\phi_i$ are the components of $\phi$. \\
Let us consider the symmetric Schr\"{o}dinger operator,
$\cL_{0}$ in $H(\G_n)$ which acts on components by
\begin{displaymath}
\cL_{0}\psi_i\equiv -\frac{d^2\psi_i}{dx_i^2}+q_i \psi_i ,
\end{displaymath}
and has domain consisting of the smooth functions 
with compact suppport in the open interval
\begin{displaymath}
D(\cL_{0})=\oplus^{n}_{i=1}C^{\infty}_{0}((0,\infty)) .
\end{displaymath}
The potentials $q_i$ are supposed to be continuous real valued functions
which are integrable with finite first moment, ie.
\begin{equation}\label{bcndxx}
\int^{\infty}_0 (1 + x) \vert q_i(x) \vert dx < \infty . 
\end{equation}
It is easy to see that the deficiency indices of $\cL_{0}$ are $(n,n)$. 
Consequently we may consider the self-adjoint extensions of $\cL_{0}$ and
indeed, using the results of Neumann extension theory \cite{Akh:Glz} 
parameterise these extensions by the unitary matrices $\U(n)$. \\
The problem of finding self-adjoint {\em boundary conditions} for
such an operator is discussed in detail in
\cite{Kost:Sch,Exn:Seb}. In \cite{Kost:Sch} all self-adjoint boundary
conditions are parameterised non-uniquely in terms of two $n$-th order
matrices, $A$ $B$, such that $(A\, B)$ is of maximal rank and
$AB^{\star}=BA^{\star}$ is hermitian (in this paper the authors consider
graphs with trivial compact part as well as graphs with
non-trivial compact part). \\
Instead, here we will use the discussion of hermitian symplectic spaces
to parameterise all of the self-adjoint boundary conditions at the origin
in terms of a unitary matrix $U$. A simple calculation using integration
by parts shows that the boundary form for the Schr\"{o}dinger operator is
\begin{equation}\label{bf1}
(\cL^{\star}_{0}\psi,\phi) - (\psi,\cL^{\star}_{0}\phi) = 
\sum^{n}_{j=1}
\left. [ \bar{\psi}_i \phi_{i,x} - \bar{\psi}_{i,x} \phi_i ]\right|_{0} . 
\end{equation}
This boundary form may be thought of as acting in the
$2n$-dimensional hermitian symplectic space, $H_{2n}$, of
boundary values at the origin. The boundary form can be written
\[
{\cal J}( \psi , \phi ) = (\psi ,\J\phi )
\] 
where on the right hand side we use the inner product in $\C^{2n}$ and 
$\psi$, $\phi$ are vectors in $\C^{2n}$ of the form
\[
( \psi_{1}|_0,\ldots ,\psi_{n}|_0, \psi_{1,x}|_0,\ldots ,\psi_{n,x}|_0
)^T .
\]
Consequently this defines a canonical basis. Let us represent the canonical
basis elements explicitly as $\{\xi_{0,i}\}^{2n}_{i=1}\in H_{2n}$ where for
$i=1,\ldots ,n$, $\xi_{0,i}$ represents the boundary condition 
$\left.\psi_{i}\right|_0 =1$; and for $i=n+1,\ldots ,2n$ it represents the
boundary condition $\left.\psi_{i,x}\right|_0 =1$. The first
$n$ and last $n$ elements of a canonical basis each span a Lagrange
plane---the first $n$ basis vectors specify self-adjoint Neumann boundary
conditions, and the last $n$ basis vectors specify self-adjoint Dirichlet
boundary conditions. \\
We fix a unitary matrix $U$ and consider the associated
self-adjoint boundary conditions specifying a Lagrange plane. From
corollary \ref{LPcorr} the basis for the Lagrange plane defined by $U$
is given by
\[
\xi_i = \sum^{2n}_{j=1} g_{ij} \xi_{0,j} \hspace{5mm} \mbox{for}\;
i=1,\ldots ,n ,
\]
where $g$ is defined by equation
(\ref{hs2}). Writing this in terms of boundary values we see that (up
to a transposition) the set of self-adjoint boundary values is
\[
(\psi_{1}|_0,\ldots ,\psi_{n}|_0, \psi_{i,x}|_0,\ldots ,\psi_{n,x}|_0 )^T
\in
\Ran \left( \begin{array}{c}
\frac{1}{2}( U+\1 ) 
\vspace{3mm} \\
\frac{i}{2}( U-\1 )
\end{array} \right) .
\]
It is convenient to have the self-adjoint boundary
{\em conditions}, ie. to have an expression in terms of the kernel
rather than the range of a matrix. This is possible if we note that 
\[
\Ran \left( \begin{array}{c}
\frac{1}{2}( U+\1 ) 
\vspace{3mm} \\
\frac{i}{2}( U-\1 )
\end{array} \right) = \ker \left( \frac{i}{2}(U^{\star}-\1), \; 
\frac{1}{2}(U^{\star}+\1) \right)
\]
which follows from equation (\ref{krii}) and the fact that both of these
matrices are of rank $n$. Consequently, the boundary conditions may be
expressed
\begin{equation}\label{basicbc}
\frac{i}{2}(U^{\star}-\1 ) \left. \psi \right|_0 +
\frac{1}{2}(U^{\star} +\1 ) \left. \psi_{x} \right|_0 = 0 .
\end{equation}

In the remainder of this subsection we will discuss how the matrix $U$,
used to describe the boundary conditions, appears in the asymptotics of
the scattering matrix. It is convenient to consider the
Schr\"{o}dinger operator on the graph with $n$ rays as a matrix
operator, with diagonal potential, see \cite{Har,Har1}. Let us consider the
matrix of $n$ solutions of Schr\"{o}dinger equation $\cL\eSw = \lambda\eSw$
on the graph satisfying the following boundary conditions at the origin
\begin{equation}\label{eSwo}
\left. \eSw \right|_0 = \frac{1}{2}( U + \1 ) \equiv A ,\qquad 
\left. \eSw_x \right|_0 = \frac{i}{2}( U -\1 ) \equiv B .
\end{equation}
It is clear, from equation (\ref{krii}), that each column of 
$\eSw$ satisfies the self-adjoint boundary conditions, ie. equation 
(\ref{basicbc}), and hence is (formally) an eigenfunction of the self-adjoint
Schr\"{o}dinger operator on the graph with boundary conditions prescribed by
$U$. \\
Likewise we can define the Jost solutions,
$\Jsd{}$, as the matrix of solutions of the homogeneous equation
$\cL\Jsd{} =\lambda\Jsd{}$, with  asymptotic behaviour
\[ 
\lim_{x\rightarrow\infty}\Jsd{}(x,k) \sim e^{\pm ikx}\1 .
\]
We denote $\lambda=k^2$. As the Jost solutions form a complete set of 
solutions we can write 
\begin{equation}\label{eSwJs}
\eSw (x,k) = \Jsb{}(x,k)\Mb{}{}(k) + \Js{}(x,k)\M{}{}(k) .
\end{equation}
In this notation we define the scattering wave solutions
\begin{eqnarray*}
\Sw (x,k) \equiv \eSw (x,k) \Mb{-1}{} = \Jsb{} + \Js{} S(k)
\end{eqnarray*}
where $S(k)$ is known as the scattering matrix. The
coefficients $\Md{}{}$ can be evaluated by taking the Wronskian of $\eSw$
and $\Js{}$ or $\Jsb{}$ \cite{Har}
\begin{equation}\label{sdmi}
\Md{}{} = \pm\frac{1}{2ik}\left[ \Jfad{} B - \Jfad{,x} A \right] .
\end{equation} 
where $\Jfd{}(k)\equiv\Jsd{}(0,k)$ are known as the Jost functions and 
$\mbox{}^{\dagger}$ is the involution
$Y^{\dagger} (x,k)\equiv Y^{\star}(x,\bar{k})$.
The Wronskian of $\eSwa$ and $\eSw$
\[
W\{\eSwa ,\eSw \} = \left. \left[ \eSwa \eSw_x - \eSwa_x \eSw \right] 
\right|_0 = A^{\star} B - B^{\star} A = 0 ,
\]
is always zero. Moreover, if we write
$\eSw$ in terms of the scattering wave solutions 
\[
W\{\eSwa ,\eSw \} = \Ma{}{} W\{\Jsa{} + S^{\dagger} \Jsab{}
,\Jsb{} + \Js{} S \} \Mb{}{} = 2ik \Ma{}{} \left[ -\1 + S^{\dagger} S
\right] \Mb{}{} ,
\]
we see, since $S^{\dagger}=S^{\star}$ for $k\in\R$, that
the scattering matrix is unitary for real $k$. \\
If we diagonalise $U$, and use the well known asympototics of the Jost
functions \cite{Agr:Mar,Har} in the above expression for $\Md{}{}$, we
see that the scattering matrix has the following asymptotic behaviour:
\begin{lem}\label{umhu}
Given the self-adjoint operator $\cL$, with associated unitary matrix $U$
defining the boundary conditions of $\cL$, the scattering matrix of $\cL$
has the asymptotics
\[
\lim_{k\rightarrow\infty} S(k) \sim \hat{U}
\]
where $\hat{U}$ is a unitary hermitian matrix derived from $U$ by applying
the map
\[
z\mapsto \left\{ \begin{array}{cc} 
1 & : z\in\T \setminus \{-1\} \\
-1 & : z = -1
\end{array} \right.
\]
to the spectrum of $U$. Here $\T$ is the unit circle in $\C$.
\end{lem}
{\it Proof:} Let us diagonalise the matrix $U$. In this basis, using
equation (\ref{sdmi}) and the asymptotics of the Jost
functions, the scattering matrix approaches
\[
\lim_{k\rightarrow\infty} - \left[ (e^{i\varphi_j} - 1 ) + k
(e^{i\varphi_j} + 1) \right]   
\left[ (e^{i\varphi_j} - 1 ) - k (e^{i\varphi_j} + 1 )\right]^{-1} 
\]
in the limit of large $k$. Here the $e^{i\varphi_j}$ are the unitary 
eigenvalues of $U$. There are two cases; when $e^{i\varphi_j}=-1$, this 
limit is $-1$, and when $e^{i\varphi_j}\neq-1$ the limit is 
$1$. \hspace*{\fill} $\Box$ \\

We note that those boundary conditions which
are defined by unitary matrices which in addition are hermitian matrices
can be expressed by projections---the terms $\frac{1}{2}(U \pm \1)$ are
really  orthogonal projections 
\[
P = \frac{1}{2}\left( U+\1\right),\qquad
P^{\perp} = \1 - P = -\frac{1}{2}\left( U-\1\right) .
\]
which follows simply from the fact that $U=U^{\star}=U^{-1}$. Using this
notation and orthogonality we can write the 
boundary conditions, equation (\ref{basicbc}), as
\begin{equation}\label{Pbc}
P^{\perp} \left. \psi \right|_0 = 0, \qquad
P \left. \psi_x \right|_0 = 0 .
\end{equation}
Consequently these boundary conditions are characterised by the fact
that the conditions on the functions and the derivatives
of the functions at the origin are independently specified. \\
The associated scattering matrix has the form
\begin{equation}\label{Psm}
S(k) = - \left[ i \Jfab{} P^{\perp} + \Jfab{,x} P \right]
         \left[ i \Jfa{}  P^{\perp} + \Jfa{,x} P \right]^{-1} .
\end{equation}
In the case of zero potential so that the Jost solutions are 
exponential functions we see that the
scattering matrix is constant
\begin{equation}\label{sh}
S(k) = - \left[ P^{\perp} - k P \right] 
\left[ P^{\perp} + k P \right]^{-1} = - P^{\perp} + P = U .
\end{equation}
Therefore the scattering wave has no poles and
there are no discrete eigenvalues. \\
In contrast if $U$ is not hermitian we will have discrete
eigenvalues, or alternatively resonances, when the potential is, apart
from at the origin, identically zero. This reproduces all cases--like
for instance a $\delta$ or $\delta^{\prime}$-interaction at the
origin---in which bound states or resonances appear for a zero-range
potential.

\section*{Acknowledgements}
The author would like to thank Prof B.S. Pavlov for his advice and
many useful conversations.


\begin{thebibliography}{10}

\bibitem{Agr:Mar}
Z.~S. Agranovich and V.~A. Marchenko.
\newblock {\em The Inverse Problem of Scattering Theory}.
\newblock Gordon and Breach, New York, 1963.

\bibitem{Akh:Glz}
N.~I. Akhiezer and I.~M. Glazman.
\newblock {\em Theory of Linear Operators in Hilbert Space}.
\newblock Frederick Ungar Publishing, New York, 1966.

\bibitem{Alb:Kur}
S.~Albeverio and P.~Kurasov.
\newblock {\em Singular Perturbations of Differential Operators}.
\newblock London Mathematical Society Lecture Note Series 271. Cambridge
  University Press, Cambridge, 2000.

\bibitem{Arn}
V.~I. Arnold.
\newblock {\em Mathematical Methods of Classical Mechanics}.
\newblock Graduate Texts in Mathematics 60. Springer, Berlin, 1978.

\bibitem{Arn:Giv}
V.~I. Arnold and A.~B. Givental.
\newblock Symplectic geometry.
\newblock In V.~I. Arnold and S.~P. Novikov, editors, {\em Dynamical Systems
  IV}. Springer, Berlin, 1990.

\bibitem{Exn:Seb}
P.~Exner and P.~{\u{S}}eba.
\newblock Free quantum motion on a branching graph.
\newblock {\em Rep. Math. Phys}, 28:7--26, 1989.

\bibitem{Fom}
A.~T. Fomenko.
\newblock {\em Symplectic Geometry}.
\newblock Gordon and Breach, 1988.

\bibitem{Har}
M.~Harmer.
\newblock {\em The Matrix {Schr\"{o}dinger} Operator and {Schr\"{o}dinger}
  Operator on Graphs}.
\newblock PhD thesis, University of Auckland, 2000.

\bibitem{Har1}
M.~Harmer.
\newblock Inverse scattering for the matrix {Schr\"{o}dinger} operator and
  {Schr\"{o}dinger} operator on graphs with general self-adjoint boundary
  conditions.
\newblock {\em ANZIAM Journal}, 43:1--8, 2002.

\bibitem{Kost:Sch}
V.~Kostrykin and R.~Schrader.
\newblock Kirchhoff's rule for quantum wires.
\newblock {\em J. Phys A: Math. Gen.}, 32:595--630, 1999.

\bibitem{Mar:Rat}
J.~E. Marsden and T.~S. Ratiu.
\newblock {\em Introduction to Mechanics and Symmetry}.
\newblock Springer, New York, 1994.

\bibitem{Nov3}
S.~P. Novikov.
\newblock Schr{\"{o}}dinger operators on graphs and symplectic geometry.
\newblock In E.~Bierstone, B.~Khesin, A.~Khovanskii, and J.~Marsden, editors,
  {\em The Arnol'dfest (Toronto, ON, 1997)}, volume~24 of {\em Fields Institute
  Communications}, pages 397--413, 1999.

\bibitem{Pav}
B.~S. Pavlov.
\newblock The theory of extensions and explicitly solvable models.
\newblock {\em Uspekhi Math. Nauk-Russian. Math. Surveys}, 42(6):127--168,
  1987.

\bibitem{Ree:Sim}
M.~Reed and B.~Simon.
\newblock {\em Methods of Modern Mathematical Physics}.
\newblock Academic Press, New York, 1972.

\end{thebibliography}
\end{document}